\documentclass[%
 reprint,
 superscriptaddress,
 amsmath,amssymb,
 aps,https://www.overleaf.com/project/5d5ae32d4618dd37df01b4f9
 pra,
floatfix,
]{revtex4-1}
\usepackage{amsmath}
\usepackage{lipsum}
\DeclareMathOperator{\sinc}{sinc}

\usepackage{graphicx}
\usepackage{dcolumn}
\usepackage{bm}
\usepackage[colorlinks,linkcolor=blue,citecolor = blue,urlcolor=blue]{hyperref}
\usepackage{booktabs}
\usepackage{float}
\usepackage{csquotes}
\usepackage{abraces}
\DeclareMathOperator{\Erf}{Erf}
\DeclareMathOperator{\arccot}{arccot}
\usepackage{commath}
\usepackage{ulem}
\usepackage{braket}
\usepackage{color}
\usepackage{cleveref}
\begin{document}

\title{Maximizing the validity of the Gaussian Approximation for the biphoton State from parametric down-conversion}

\author{Baghdasar Baghdasaryan}
\email{baghdasar.baghdasaryan@uni-jena.de}
\affiliation{Theoretisch-Physikalisches Institut, Friedrich-Schiller-Universit\"at Jena, 07743 Jena, Germany}
\affiliation{Helmholtz-Institut Jena, 07743 Jena, Germany}


\author{Fabian Steinlechner}
\email{fabian.steinlechner@uni-jena.de}
 \affiliation{Fraunhofer Institute for Applied Optics and Precision Engineering IOF, 07745 Jena, Germany}
 \affiliation{Abbe Center of Photonics, Friedrich-Schiller-University Jena, 07745 Jena, Germany}

\author{Stephan Fritzsche}
\affiliation{Theoretisch-Physikalisches Institut, Friedrich-Schiller-Universit\"at Jena, 07743 Jena, Germany}
\affiliation{Helmholtz-Institut Jena, 07743 Jena, Germany}
 \affiliation{Abbe Center of Photonics, Friedrich-Schiller-University Jena, 07745 Jena, Germany}
\date{\today}

\begin{abstract}
Spontaneous parametric down-conversion (SPDC) is widely used in quantum applications based on photonic entanglement. The efficiency of photon pair generation is often characterized by means of a $\sinc(L\Delta k/2)$ function, where $L$ is the length of the nonlinear medium and $\Delta k$ is the phase mismatch between the pump and down-converted fields. In theoretical investigations, the \textit{sinc} behavior of the phase mismatch has often been approximated by a Gaussian function $\exp{(-\alpha x^2)}$ in order to derive analytical expressions for the SPDC process. Different values have been chosen in the literature for the optimization factor $\alpha$, for instance, by comparing the widths of \textit{sinc} and Gaussian functions or the momentum of down-converted photons. As a consequence, different values of $\alpha$ provide different theoretical predictions for the same setup. Therefore an informed and unique choice of this parameter is necessary. In this paper, we present a choice of $\alpha$ which maximizes the validity of the Gaussian approximation. Moreover, we also discuss the so-called \textit{super}-Gaussian and \textit{cosine}-Gaussian approximations as practical alternatives with improved predictive power for experiments. 
\end{abstract}

\pacs{Valid PACS appear here}

\maketitle

\section{Introduction}
In spontaneous parametric down-conversion (SPDC), a nonlinear-responding quadratic crystal is pumped by a laser field, in order to convert (high-energy) photons into entangled photon pairs. Entangled states generated by SPDC have provided an experimental platform for fundamental quantum technologies, such as quantum cryptography \cite{cryptography}, quantum teleportation \cite{teleportation}, or optical quantum information processing \cite{RevModPhys.79.135}, and by including recent milestone experiments in photonic quantum computing \cite{doi:10.1126/science.abe8770}.

The quantum theory of down-converted pairs is well known and plays a crucial role in the modeling of SPDC experiments. The state of such a SPDC pair is also known as a biphoton state. Within the paraxial approximation, where the longitudinal and transversal components of the wave vector are treated separately, $\bm{k}=\bm{q}+k_z(\omega)\bm{z}$, the biphoton state can be written as \cite{WALBORN201087,Karan_2020,https://doi.org/10.48550/arxiv.2208.09423}
\begin{align}\label{SPDCstate}
    \ket{\Psi} =N \iint & d\bm{q}_s \: d\bm{q}_i \:d\omega_s \: d\omega_i\: \overbrace{\mathrm{V}(\bm{q}_s+\bm{q}_i)\,\mathrm{S}_p(\omega_s+\omega_i)}^{pump} \nonumber\\&
 \times \underbrace{ \sinc\biggl(\frac{L\Delta k_z}{2}\biggr)}_{phase \medspace matching}\,
\hat{a}^{\dagger}_s(\bm{q}_s,\omega_s)\:\hat{a}^{\dagger}_i(\bm{q}_i,\omega_i)\ket{vac}.
\end{align}
In expression \eqref{SPDCstate}, $N$ is the normalization factor, $V(\bm{q}_p)$ is the spatial and $S_p(\omega_p)$ the spectral distribution of the pump beam, $L$ is the length of the nonlinear crystal, $\ket{vac}$ is the vacuum state, $\omega_{s,i}$ and $\bm{q}_{s,i}$ are the energies and transverse components of wave vectors of down-converted fields (signal and idler), and $\hat{a}^{\dagger}_{s,i}(\bm{q}_{s,i})$ are the corresponding creation operators. 

Expression \eqref{SPDCstate} can be simplified if we approximate the \textit{sinc} function in terms of the Gaussian function $\sinc(x^2)\approx \exp(-x^2)$. Many useful analytical expressions can be then defined for SPDC within this approximation. A good example is the expression for the Schmidt number of the biphoton state presented in Ref. \cite{PhysRevLett.92.127903}.

Already in Ref. \cite{PhysRevLett.92.127903}, the calculations within the Gaussian approximation delivered a small deviation from experimentally measured values for the Schmidt number. It has often been suggested in the literature to optimize this approximation by presenting an optimization factor $\alpha$ in the exponential expression $\exp{(-\alpha x^2)}$. The value of $\alpha$ has been chosen, for example, by matching the widths of Gaussian and \textit{sinc} functions \cite{PhysRevA.53.4360,PhysRevLett.99.243601,Miatto2012}, by matching second-order momenta \cite{Schneeloch_2016}, by comparing momentum
correlations in SPDC \cite{Gomez:12}, or by comparing the coincidence and single-particle spectral widths of the biphoton state \cite{Fedorov_2009}.

Obviously, different values of $\alpha$ deliver different theoretical predictions for the same setup. A unique choice of $\alpha$, which should minimize the error of the approximation, is lacking. In order to solve this problem, we look in this paper at the states themselves, instead of comparing just the $\textit{sinc}$ and Gaussian functions or other observable. Eventually, the goal is to make the distance between $\textit{sinc}$- and Gaussian-like states as small as possible. We use an appropriate distance measure for this purpose such as fidelity \cite{Nielsen} and find the particular $\alpha$ that maximizes it. Except for the Gaussian approximation, we will also discuss the \textit{cosine}- and \textit{super}-Gaussian approximations and will compare them with the usual Gaussian approximation. (All these functions can be compared in Fig. \ref{fig3}, where we already included the optimized values of $\alpha$.)

\section{Theory and results}
We first should determine the pump and phase-matching characteristics from expression \eqref{SPDCstate}, in order to calculate the fidelity of the \textit{sinc}- and Gaussian-like states. The pump beam is usually fixed by the experimental setup, and its function in expression \eqref{SPDCstate} is well known. The phase mismatch in the $z$ direction $\Delta k_z=k_{p,z}-k_{s,z}-k_{i,z}$ requires more careful determination, which depends on many characteristics of the experiment, such as the crystal type, the polarization of interacting beams, and the geometry of the setup. The derivation of $\Delta k_z$ has already been reported for a very general experimental scenario in Ref. \cite{https://doi.org/10.48550/arxiv.2208.09423}. In comparison to Ref. \cite{https://doi.org/10.48550/arxiv.2208.09423}, here, we additionally assume degenerate momentum vectors for signal and idler photons $k_p\approx2k_s$, which allows us to apply the Gaussian approximation. With this in mind, the phase mismatch $\Delta k_z$ can be written as \cite{https://doi.org/10.48550/arxiv.2208.09423}
%
%
%
\begin{equation}\label{phaseMatching}
\Delta k_z= \frac{\Omega_s+\Omega_i}{u_p}-\frac{\Omega_s}{u_s}-\frac{\Omega_i}{u_i}+\frac{\abs{{\bm{q}_s-\bm{q}_{i}}}^2}{2k_p},
\end{equation}
where $\Omega_j$ is the deviation from the central frequency $\omega_{j,0}$, $\omega_j=\omega_{j,0}+\Omega_j$ with the assumption $\Omega_j\ll\omega_{j,0}$, and $u_j=1/(\partial k_j/\partial \Omega)$ is the group velocity evaluated at the central frequency.

The right-hand side of Eq. \eqref{phaseMatching} can be divided into two parts. The first three terms represent the spectral and the last term the spatial properties of the biphoton state. The spectral and spatial degrees of freedom (DOFs) are in general coupled due to the phase-matching characteristics in SPDC \cite{Osorio_2008,PhysRevLett.102.223601,https://doi.org/10.48550/arxiv.2208.09423,sevillagutirrez2022spectral}. However, under certain approximations, such as the narrowband \cite{PhysRevA.83.033816}, thin-crystal \cite{Yao_2011,PhysRevA.103.063508}, or plane wave approximations \cite{PhysRevLett.99.243601}, either only the first three terms survive (frequency-resolved biphoton state) or only the last one survives (spatially resolved biphoton state). In this paper, we first develop the Gaussian approximation for spectral and spatial DOFs separately. The description of the coupling in spectral and spatial domains is more challenging in the scope of the Gaussian approximation, which is also discussed in this paper. 

\subsection{Spatially resolved biphoton state}
The signal and idler fields can be assumed to be monochromatic if narrowband filters are used in front of the detectors. This is called the narrowband approximation, which transforms the biphoton state into
 \begin{align}\label{sinc}
    \ket{\Psi} =N \iint & d\bm{q}_s \: d\bm{q}_i \mathrm{V}(\bm{q}_s+\bm{q}_i)\:  \sinc\biggl(\frac{L\abs{{\bm{q}_s-\bm{q}_{i}}}^2}{4k_p}\biggr) \nonumber\\&
 \times \hat{a}^{\dagger}_s(\bm{q}_s)\:\hat{a}^{\dagger}_i(\bm{q}_i)\ket{vac}.
\end{align}
Expression \eqref{sinc} has now a simple form so that we can apply the Gaussian, \textit{super}-Gaussian, or \textit{cosine}-Gaussian approximations and calculate the corresponding fidelity. We will present the detailed calculation only for the Gaussian approximation since the derivations for the \textit{super}-Gaussian and \textit{cosine}-Gaussian approximations are similar.
\subsubsection{Gaussian approximation} 
Similar to expression \eqref{SPDCstate}, the approximated Gaussianstate is written as
 \begin{align}\label{gussian}
    \ket{\Psi_G} =N_G \iint & d\bm{q}_s \: d\bm{q}_i \mathrm{V}(\bm{q}_s+\bm{q}_i)\:  \exp\biggl(-\alpha\frac{L\abs{{\bm{q}_s-\bm{q}_{i}}}^2}{4k_p}\biggr) \nonumber\\&
 \times \hat{a}^{\dagger}_s(\bm{q}_s)\:\hat{a}^{\dagger}_i(\bm{q}_i)\ket{vac},
\end{align}
where $N_G$ is the new normalization constant. The fidelity of the states from Eqs. \eqref{sinc} and \eqref{gussian}, which is simply the scalar product of these states, is then given by
  \begin{align}\label{scalar}
    \braket{\Psi|\Psi_G} =&N\cdot N_G \iint d\bm{q}_s \: d\bm{q}_i \,\abs{\mathrm{V}(\bm{q}_s+\bm{q}_i)}^2\nonumber\\& \times\sinc\biggl(\frac{L\abs{{\bm{q}_s-\bm{q}_{i}}}^2}{4k_p}\biggr)\: \exp\biggl(-\alpha\frac{L\abs{{\bm{q}_s-\bm{q}_{i}}}^2}{4k_p}\biggr).
\end{align}
The integral is difficult to calculate in the momentum space but mathematically more straightforward in the position space. We use the notations $\bm{q}_{-}=\bm{q}_s-\bm{q}_i$ and $\bm{q}_{+}=\bm{q}_s+\bm{q}_i$ and rewrite the function from Eq. \eqref{scalar} with their Fourier transforms 
\begin{align*}
    \braket{\Psi|\Psi_G}=& N\cdot N_G \iint d\bm{q}_s \: d\bm{q}_i\, \frac{1}{2 \pi}\int d\bm{r}\: \Phi_F(\bm{r})\,e^{i\bm{r}\bm{q}_{-}}\\&\nonumber
   \times \frac{1}{2 \pi}\int d\bm{r}^{'}\, \mathrm{V}_F(\bm{r}^{'})\,e^{i\bm{r}^{'}\bm{q}_{+}},
\end{align*}
where the transformed functions are given by
\begin{align}\label{fourier}
   \Phi_F(\bm{r})=&\frac{1}{2 \pi} \int d\bm{q}_{-}\, \sinc\biggl(\frac{L\abs{{\bm{q}_{-}}}^2}{4k_p}\biggr)\: \exp\biggl(-\alpha\frac{L\abs{{\bm{q}_{-}}}^2}{4k_p}\biggr)\nonumber\\& \times e^{-i\bm{r}\bm{q}_{-}}
\end{align}
and
\begin{equation}\label{fourier2}
  \mathrm{V}_F(\bm{r}^{'})=\frac{1}{2 \pi} \int d\bm{q}_{+} \abs{\mathrm{V}(\bm{q}_{+})}^2 \,e^{-i\bm{r}^{'}\bm{q}_{+}}.
\end{equation}
The integrals over momentum space can now be implemented
\begin{align*}
    \braket{\Psi|\Psi_G}=& N\cdot N_G \iint d\bm{r}\: d\bm{r}^{'} \:\Phi_F(\bm{r})\,\mathrm{V}_F(\bm{r}^{'})\\&\nonumber
   \times \frac{1}{2 \pi}\int d\bm{q}_s\, \,e^{i\bm{q}_s(\bm{r}+\bm{r}^{'})} \frac{1}{2 \pi}\int d\bm{q}_i \,e^{-i\bm{q}_i(\bm{r}-\bm{r}^{'})},
\end{align*}
that give rise to two \textit{delta} functions
\begin{align*}
    \braket{\Psi|\Psi_G}=& N\cdot N_G \iint  d\bm{r}\: d\bm{r}^{'} \:\Phi_F(\bm{r})\,\mathrm{V}_F(\bm{r}^{'})\\&\nonumber
   \times  (2\pi)^2\,\delta{(\bm{r}+\bm{r}^{'})}  \,\delta{(\bm{r}-\bm{r}^{'})}.
\end{align*}
The integrals over delta functions are trivial to perform, which gives us:
\begin{equation}\label{scalarG}
        \braket{\Psi|\Psi_G} =N\cdot N_G\:\Phi_F(0)\,\mathrm{V}_F(0) \pi^2.
\end{equation}
Note that expression \eqref{scalarG} has been derived independent of the pump beam and is quite universal. Moreover, we can see from Eq. \eqref{fourier2} that $\mathrm{V}_F(0)$ is equal just to $1/2 \pi$ if the pump field $V(\bm{q}_p)$ is normalized, i.e., the fidelity is independent of the spatial distribution of the pump beam. 

\begin{figure}[t!]
\includegraphics[width=.47\textwidth]{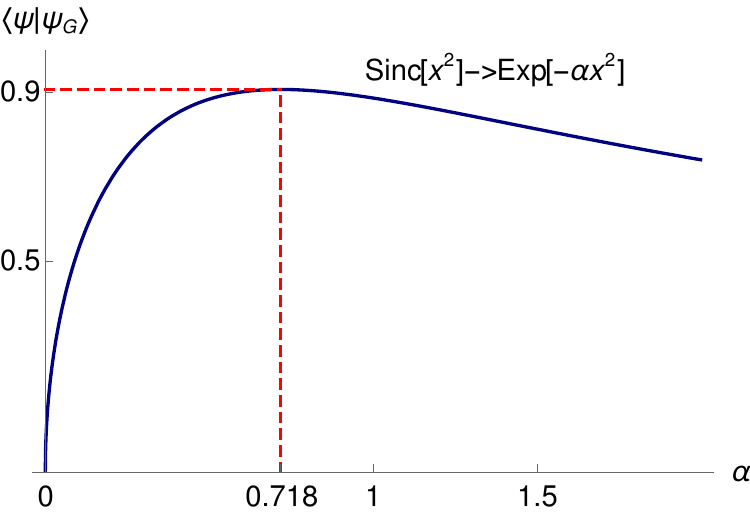}
\caption{Dependence of the fidelity of the \textit{sinc}- and Gaussian-like states on the optimization factor $\alpha$. The maximum possible overlap is equal to $0.9$ for $\alpha=0.718.$}\label{fig1}
\end{figure}

 %
%
The normalization constants from \eqref{scalarG} can be calculated in the same way by using the \textit{method} of the Fourier transform [Eq. \eqref{scalarG} for the case  $\braket{\Psi|\Psi}$ or $ \braket{\Psi_G|\Psi_G}$], giving $N=\sqrt{2L/(k_p\pi^2)}$ and $N_G=\sqrt{2L\alpha/(k_p\pi)}$, respectively. In the last step, we calculate the missing function value $\Phi_F(0)$ of Eq. \eqref{scalarG} from Eq. \eqref{fourier} and come up with the final expression for the fidelity
\begin{equation}\label{finalG}
\braket{\Psi|\Psi_G}=2\sqrt{\frac{\alpha}{\pi}}\arccot{\alpha}.
\end{equation}
The first thing to be remarked is that the fidelity is independent of the pump and crystal characteristics and depends only on the optimization factor $\alpha$. Let us now find out the maxima of this function. Expression \eqref{finalG} as a function of $\alpha$ is presented in Fig. \ref{fig1}. The maximum is reached for $\alpha=0.718$ with the fidelity value $\braket{\Psi|\Psi_G}_{max}=0.9$. Note that this is very close to the value $\alpha=0.724$ such that the $\textit{sinc}$ and Gaussian functions have equal widths at $1/e^2$ from the peak intensity. 

Thus the Gaussian approximation delivers the best matching with the $\textit{sinc}$-like state if $\alpha=0.718$ giving an overlap value of $0.9$.
\subsubsection{Super-Gaussian} 
\begin{figure}[t!]
\includegraphics[width=.47\textwidth]{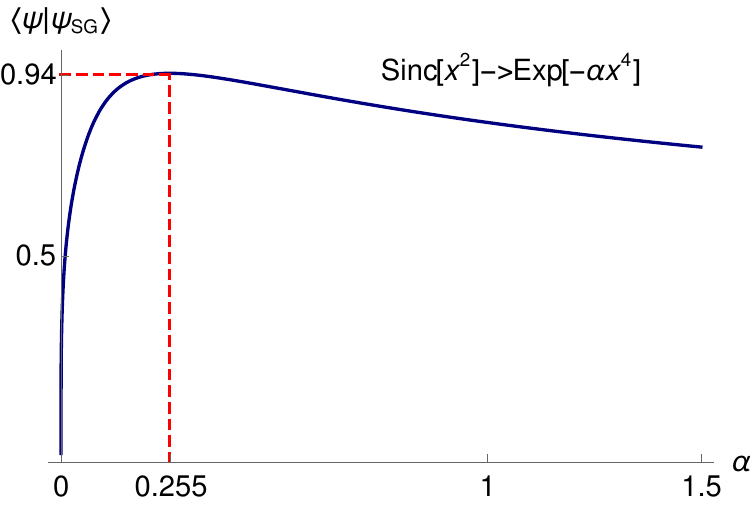}
\caption{The same as in Fig. \ref{fig1} but for the \textit{super}-Gaussian-like state. The maximum possible overlap is equal to $0.94$ for $\alpha=0.255.$}\label{fig2}
\end{figure}

The \textit{sinc} function can also be approximated to a \textit{super}-Gaussian function \cite{Miatto2012}
 \begin{align}
    \sinc{\biggl(\frac{L\abs{{\bm{q}_s-\bm{q}_{i}}}^2}{4k_p}\biggr)}\approx \exp\biggl[-\alpha\biggl(\frac{L\abs{{\bm{q}_s-\bm{q}_{i}}}^2}{4k_p}\biggr)^2\biggr] .
\end{align}
The derivation of the scalar product $\braket{\Psi|\Psi_{SG}}$ can be performed in the same way as for the Gaussian approximation. The normalization constant for the state with a \textit{super}-Gaussian phase matching is given by $N_{SG}=\sqrt{2\alpha/\pi}\,2L/(k_p\pi)$. The corresponding fidelity reads as
\begin{equation}\label{finalSG}
    \braket{\Psi|\Psi_{SG}}=(2\pi \alpha)^{0.25}\Erf\biggl[{\frac{1}{2\sqrt{\alpha}}}\biggr],
\end{equation}
which like expression \eqref{finalG}, is independent of the pump and crystal characteristics. The fidelity \eqref{finalSG} as a function of $\alpha$ is presented in Fig. \ref{fig2}. The  maximum overlap value $\braket{\Psi|\Psi_G}_{max}\approx 0.943$ can be achieved if the optimization factor is $\alpha=0.255$. Although the fidelity $\braket{\Psi|\Psi_{SG}}$ can be higher than that of the Gaussian approximation, the \textit{super}-Gaussian approximation is not advantageous from a mathematical point of view, since the term $\abs{{\bm{q}_s-\bm{q}_{i}}}^4$ with the fourth power of transverse momenta appears in the exponential expression and makes it difficult to derive analytical expressions for SPDC.
\subsubsection{Cosine-Gaussian approximation} As suggested in Ref.\cite{PhysRevA.80.013811}, we can consider, instead of a Gaussian, a more general \textit{cosine}-Gaussian approximation
 \begin{align}
    \sinc{\biggl(\frac{L\abs{{\bm{q}_s-\bm{q}_{i}}}^2}{4k_p}\biggr)}\approx & \exp\biggl(-\alpha\frac{L\abs{{\bm{q}_s-\bm{q}_{i}}}^2}{4k_p}\biggr)\nonumber\\&
    \times \cos\biggl(\beta\frac{L\abs{{\bm{q}_s-\bm{q}_{i}}}^2}{4k_p}\biggr).\label{cos}
\end{align}
This is a better approximation than the usual Gaussian approximation, since the side oscillation of the \textit{sinc} function can be optimized with the $\textit{cosine}$ function. Moreover, the \textit{cosine} function can be rewritten by using Euler's formula
 \begin{align}\label{cosine}
    \sinc{\biggl(\frac{L\abs{{\bm{q}_s-\bm{q}_{i}}}^2}{4k_p}\biggr)}\approx Re\biggl\{  \exp\biggl[-(\alpha-i\beta)\frac{L\abs{{\bm{q}_s-\bm{q}_{i}}}^2}{4k_p}\biggr]\biggr\},
\end{align}
which brings us back to the simple Gaussian-like function with the optimization factor $\alpha-i\beta$. Therefore all formulas derived for the Gaussian approximation can be easily translated into the \textit{cosine}-Gaussian case by replacing the optimization factors $\alpha\to\alpha-i\beta$ and taking then the real part of the derived expression. The normalization constant of the state with phase matching \eqref{cosine} is
\begin{equation*}
  N_{CG}= \sqrt{ \frac{4L}{\pi k_p}\,}\sqrt{\frac{\alpha^3+\alpha \beta^2}{2\alpha^2+\beta^2}},
\end{equation*}
and consequently, the overlap integral takes the form
\begin{align*}
     \braket{\Psi|\Psi_{CG}}=&\sqrt{\frac{2}{\pi}}\biggr(\arctan{\frac{1-\beta}{\alpha}}+\arctan{\frac{1+\beta}{\alpha}}\biggl)\\&
     \times \sqrt{\frac{\alpha^3+\alpha \beta^2}{2\alpha^2+\beta^2}}.
\end{align*}
We calculated the maximal fidelity to be $\braket{\Psi|\Psi_{CG}}_{max}\approx 0.94$ for the optimization factors $\alpha=0.39$ and $\beta =0.49$. Thus the same fidelity of the \textit{super}-Gaussian approximation can be achieved also for the \textit{cosine}-Gaussian approximation by keeping the Gaussian-like behavior.

\begin{figure}[t!]
\includegraphics[width=.47\textwidth]{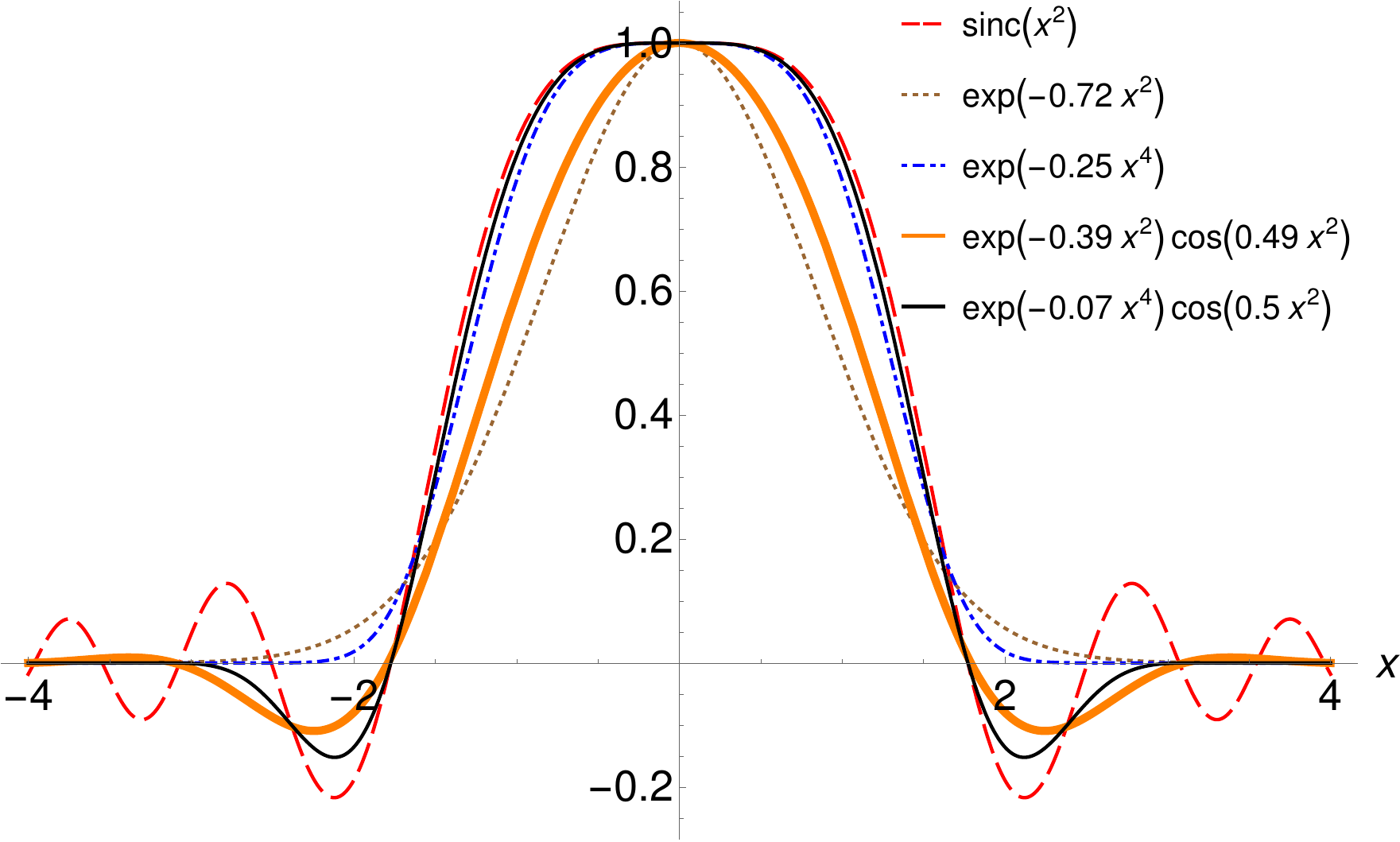}
\caption{Comparison of the \textit{sinc} function with approximated functions. As expected, \textit{cosine}-Gaussian-like functions also approximate the oscillations of the \textit{sinc} function and deliver a better matching compared with a simple Gaussian function.}\label{fig3}
\end{figure}

In Table \ref{table1}, we summarized the approximations and the corresponding optimization factors for the spatially-resolved biphoton state. In addition, we also presented the \textit{cosine}-\textit{super}-Gaussian approximations for the spatially resolved biphoton state that delivers the highest fidelity $0.97$. However, again from a mathematical point of view, it is not convenient to deal with a \textit{super}-Gaussian function, because of the term $\abs{{\bm{q}_s-\bm{q}_{i}}}^4$ in the exponential expression. The functions themselves are presented in Fig. \ref{fig3}.
  \begin{table*}
\begin{ruledtabular}
\begin{tabular}{ccccc}
  $\sinc{x^2}\approx$ & $\exp{(-\alpha x^2)}$&$\exp{(-\alpha x^4)}$&$\exp{(-\alpha x^2)}\cos{\beta x^2}$ &$\exp{(-\alpha x^4)}\cos{\beta x^2
 }$\\ \hline
 $\alpha,\beta$ &$0.72$&$0.25$ &$0.39,\:0.49$&$0.07,\:0.5$ \\
Fidelity&$0.9$
 &$0.94$&$0.94$&$0.97$\\
\end{tabular}
\end{ruledtabular}\caption{Optimization factors of Gaussian, \textit{super}-Gaussian, \textit{cosine}-Gaussian, and \textit{cosine}-\textit{super}-Gaussian approximations and the corresponding fidelities.}\label{table1}
\end{table*}
\subsection{Frequency-resolved biphoton state}\label{spectral}
We assume large area collection modes $\bm{q_s}=\bm{q_i}\approx0$ \cite{PhysRevLett.99.243601}, in order to remain only with the spectral part of the right-hand side of Eq. \eqref{phaseMatching}. The biphoton state transforms into
 \begin{align}\label{spectralState}
    \ket{\Psi} =N \iint & d\Omega_s \: d\Omega_i\, \mathrm{S}_p(\Omega_{s}+\Omega_i)\:  \sinc\biggl(\frac{L\Delta k_z}{2}\biggr) \nonumber\\&
 \times \hat{a}^{\dagger}_s(\Omega_s)\:\hat{a}^{\dagger}_i(\Omega_i)\ket{vac},
\end{align}
where the phase mismatch now is
\begin{equation*}
    \Delta k_z =\frac{\Omega_s+\Omega_i}{u_p}-\frac{\Omega_s}{u_s}-\frac{\Omega_i}{u_i}.
\end{equation*}
The temporal distribution of the pump with a Gaussian envelope of pulse duration $t_0$ can be modeled as $\mathrm{S}_p(\omega_p)=\exp{[-(\omega_p-\omega_{p,0})^2\,t_0^2/4]}$. The corresponding Gaussian-like state is given by
 \begin{align}\label{frequencyS}
    \ket{\Psi_G} =N \iint & d\Omega_s \: d\Omega_i\, \mathrm{S}_p(\Omega_{s}+\Omega_i)\:  \exp\biggl[-\alpha\biggl(\frac{L\Delta k_z}{2}\biggr)^2\biggl] \nonumber\\&
 \times \hat{a}^{\dagger}_s(\Omega_s)\:\hat{a}^{\dagger}_i(\Omega_i)\ket{vac}.
\end{align}
Unfortunately, we could not find any analytical expression for the fidelity $\braket{\Psi|\Psi_G}$ and performed the calculations numerically. The optimum value of $\alpha$ is $0.25$, similar to the \textit{super}-Gaussian approximation of the spatially resolved biphoton state. Although the value $\alpha=0.25$ is independent of $t_0$, the maximum value for the fidelity $\braket{\Psi|\Psi_G}_{max}$ does depend on the pulse duration, but not on the crystal length $L$. So, if $t_0$ is large enough so that the pump spectral envelope $\mathrm{S}_p(\omega_p)$ dominates over the \textit{sinc} function, the maximum overlap value is close to unity. This means that the \textit{sinc} term is negligible. If the \textit{sinc} function starts to dominate over $\mathrm{S}_p(\omega_p)$ (small $t_0$), the fidelity decreases till the lowest value $0.95$, i.e., $\braket{\Psi|\Psi_G}_{max}(t_0)\geq 0.95$, for $\alpha=0.25$.

Up to this point, we handled the Gaussian-like state \eqref{frequencyS} as an approximation to the real \textit{sinc}-like state \eqref{spectralState}. However, in recent years, domain-engineered crystals with a Gaussian phase-matching function have attracted attention. These engineered crystals have been mainly used in order to engineer pure spectral states \cite{Tambasco:16,Graffitti_2017,Graffitti:18}, where only the spectral DOF of the biphoton state has been considered. Thus the state from Eq. \eqref{frequencyS} can be easily generated in experiments. On the other hand, a general investigation of domain-engineering techniques for the spatiotemporal state is still lacking and could be an interesting topic to look at, for instance, in the scope of generating a heralded single spatiotemporal mode.
\subsection{Spatiotemporal biphoton state}\label{spationspectral}
In this section, we investigate, how well the Gaussian approximation preserves the coupling between the spatial and spectral DOFs in SPDC. The description of the spatiotemporal biphoton state with a Gaussian phase-matching function is rather rare but has been successfully used to characterize the correlation strength between these two DOFs \cite{Osorio_2008}. Here, the phase-matching function $\sinc(L\Delta k_z/2)$ is again approximated by $\exp{[-\alpha(L\Delta k_z/2})^2]$, but the momentum mismatch is characterized by the full expression from Eq. \eqref{phaseMatching}. Surprisingly, this approximation holds very well for $\alpha=0.25$ giving a fidelity value of $0.94$. Similar to the frequency-resolved biphoton state, the fidelity depends on the pulse duration $t_0$, but not on the crystal length. The argument here is the same: As long as the \textit{sinc} function dominates over the pump function, the fidelity remains the same with a value of $0.94$. This is the minimum value for the fidelity, and it becomes larger if $t_0$ is increased.
\subsection{Biphoton state in the Laguerre-Gaussian basis} At the end of this paper, we apply the Gaussian approximation to the results from our previous work. In Ref. \cite{https://doi.org/10.48550/arxiv.2208.09423}, we derived a general expression for the spatiotemporal biphoton state and successfully tested it in Ref. \cite{sevillagutirrez2022spectral}. The expression for the biphoton state from Ref. \cite{https://doi.org/10.48550/arxiv.2208.09423} still includes one-dimensional integration over the crystal length, which should be performed numerically. The Gaussian or \textit{cosine}-Gaussian approximations would transform that expression into a fully analytical form.

We construct the Gaussian approximation for the spatially-resolved biphoton state by ignoring the dispersive properties of interacting beams, so that the phase matching has the form $\abs{{\bm{q}_s-\bm{q}_{i}}}^2/2k_p$. We consider the biphoton state with a Laguerre-Gaussian (LG) pump beam of well-defined radial index $p$ and projection of orbital angular momentum (OAM) $\ell$ 
\begin{align}
    \ket{\Psi_{p,\ell}} =N \iint & d\bm{q}_s \: d\bm{q}_i \: \mathrm{LG}_{p}^{\ell}(\bm{q}_s+\bm{q}_i)\,\exp\biggl(-\alpha\frac{L\abs{{\bm{q}_s-\bm{q}_{i}}}^2}{4k_p}\biggr) \nonumber\\& \times
  \hat{a}^{\dagger}_s(\bm{q}_s)\:\hat{a}^{\dagger}_i(\bm{q}_i)\ket{vac},
\end{align}
where the spatial distribution of the pump in the momentum space is given by
\begin{align} \label{LG}
  \mathrm{LG}_{p}^{\ell}(\rho,\varphi)
    =&e^{\frac{-\rho^2\,w^2}{4}}\,e^{i\ell\,\varphi}\,\sum_{u=0}^p\, T_u^{p,\ell}\, \rho^{2u+\abs{\ell}}
\end{align} 
with $ T_u^{p,\ell}$ being
\begin{align*}
   T_u^{p,\ell}= &\sqrt{\frac{p!\,(p+|\ell|)!}{\pi}}\,
   \biggr(\frac{ w}{\sqrt{2}}\biggl)^{2u+|\ell|+1}\,\frac{(-1)^{p+u}(i)^{\ell}}{(p-u)!\,(\abs{\ell}+u)!\,u!}
.\end{align*}
Similar to Ref. \cite{https://doi.org/10.48550/arxiv.2208.09423}, we perform a mode decomposition of the biphoton state in the LG basis $\ket{p,\ell}=\int  d\bm{q}\, \mathrm{LG}_{p}^{\ell}(\bm{q})\,  \hat{a}^{\dagger}(\bm{q}) \ket{vac} $ for the characterization of the biphoton spatial DOF
\begin{eqnarray}\label{decomposition}
    \ket{\Psi_{p,\ell}}
    & = & \sum_{p_s,p_i=0}^{\infty} \sum^{\infty}_{\ell_s,\ell_i=-\infty}C_{p,p_s,p_i}^{\ell,\ell_s,\ell_i} \ket{p_s,\ell_s}\ket{p_i,\ell_i},
\end{eqnarray}
where the coincidence amplitudes can be calculated from the overlap integral $ C^{\ell,\ell_s,\ell_i}_{p,p_s,p_i} = \braket{p_s,\ell_s;p_i,\ell_i |\Psi_{p,\ell}}$,
\begin{align}\label{coe1}
    C^{\ell,\ell_s,\ell_i}_{p,p_s,p_i} 
    =   N \iint  d\bm{q}_s \: d\bm{q}_i \:  \mathrm{LG}_{p}^{\ell}(\bm{q}_s+\bm{q}_i)\,\exp\biggl(-\alpha\frac{L\abs{{\bm{q}_s-\bm{q}_{i}}}^2}{4k_p}\biggr)\nonumber &\\
        \times[\mathrm{LG}_{p_s}^{\ell_s}(\bm{q}_s)]^*  [\mathrm{LG}_{p_i}^{\ell_i}(\bm{q}_i)]^*,
\end{align}
and where the parameters \{$p_s,p_i$\} and \{$\ell_s,\ell_i$\} are radial and OAM mode numbers of signal and idler photons, respectively.
The derivation of $C^{\ell,\ell_s,\ell_i}_{p,p_s,p_i}$ is similar to the derivation from Ref. \cite{https://doi.org/10.48550/arxiv.2208.09423}; hence we show only the final expression that for $\ell \geq 0$ reads as
\begin{align} \label{expression}
    C_{p,p_s,p_i}^{\ell,\ell_s,\ell_i}
     = & N\,\pi^2\: \delta_{\ell,\ell_s+\ell_i} \nonumber\\
   &   \sum_{u=0}^{p}\sum_{s=0}^{p_s}\sum_{i=0}^{p_i} T_u^{p,\ell}\: (T_s^{p_s,\ell_s})^* \:(T_i^{p_i,\ell_i})^*\: \sum_{n=0}^{\abs{\ell}}\sum_{m=0}^{u}\nonumber\\
   &  
  \binom{\abs{\ell}}{n}\binom{u}{m}\sum_{f=0}^{u-m}\sum_{v=0}^{m}\: \binom{u-m}{f}\binom{m}{v} \Gamma[h]\:\Gamma[b]\nonumber\\
   &  
\frac{D^{d}}{H^{h}\: B^{b}}\: {_2}{\Tilde{F}}_1\biggl[h,b, 1+d,\frac{D^2}{H \,B
  }\biggl],
\end{align} 
and $ C_{p,p_s,p_i}^{\ell,\ell_s,\ell_i}= (C_{p,p_s,p_i}^{-\ell,-\ell_s,-\ell_i})^*$ for $\ell \leq 0$. The function ${_2}{\Tilde{F}}_1$ is known as the \textit{regularized} \textit{hypergeometric} function \cite{Hypergeometric2F1}, and the missing coefficients in expression \eqref{expression} are given by
\begin{eqnarray*}
 H &=& \frac{w_p^2}{4}+\frac{w_s^2}{4}+\frac{\alpha L}{4k_p}, \qquad  D = -\frac{ w_p^2}{4}+\frac{\alpha L}{4k_p}               \\[0.1cm]
  B& = &  \frac{w_p^2}{4}+\frac{w_i^2}{4}+\frac{\alpha L}{4k_p}, \qquad d =\ell_i+m-n-2v.  \\[0.1cm]
  h & = & \frac{1}{2}(2+2s+\ell+\ell_i+2(-f+u)-2n-2v+\abs{\ell_s}), \\[0.1cm]
  b &=& \frac{1}{2}(2+2f+2i+\ell_i+2m-2v+\abs{\ell_i}).
\end{eqnarray*}
Expression \eqref{expression} is now fully analytical. We can then plug in either $\alpha=0.72$ with corresponding fidelity of $0.9$ or $\alpha=0.39-0.49 i$ for the \textit{cosine}-Gaussian approximation with fidelity of $0.94$ by then taking the real part of expression \eqref{expression}, $\mathrm{Re}[C_{p,p_s,p_i}^{\ell,\ell_s,\ell_i}]$. This expression can be used to model experiments involving very high spatial modes, where the numerical approach could fail or take too much computational power.

\section{Conclusion}
The Gaussian-like phase-matching functions turn out to deliver very good approximations to \textit{sinc}-like states. For most experiments, we recommend the \textit{cosine}-Gaussian approximation with $\alpha=0.39$ and $\beta=0.49$ for a spatially resolved biphoton state, as this provides analytical results for many parameters of interest with a high fidelity. For a spatiotemporal and frequency-resolved biphoton state, we suggest using the Gaussian approximation with $\alpha=0.25$. However, with more challenging experiments aiming to maximize fiber coupling efficiencies or spatial entanglement, we still recommend using the general \textit{sinc}-like state.
\bibliographystyle{apsrev4-1}
\bibliography{bibliography.bib}      
\end{document}